\documentclass[aps,prl,twocolumn,amsmath]{revtex4}
\usepackage{epsfig}
\usepackage{amssymb}
\usepackage{color}
\definecolor{cream}{rgb}{1,1,0.7}
\definecolor{orange}{rgb}{1,0.644,0}

\newcommand{\detail}[1]{\begin{center}\fcolorbox{orange}{cream}{\begin{minipage}{0.9\hsize}\small #1 \end{minipage}}\end{center}}
\newcommand{\He}{$^3$He}
\newcommand{\Hefour}{$^4$He}

\newcommand{\taud}{\tau_{\text{\tiny D}}}
\newcommand{\diff}{\text{D}_{\text{\tiny M}}}

\def\vj{{\bf j}}

\def\vr{{\bf r}}
\newcommand{\grad}{\mbox{\boldmath$\nabla$}}
\newcommand{\be}{\begin{equation}}
\newcommand{\ee}{\end{equation}}
\newcommand{\ber}{\begin{eqnarray}}
\newcommand{\eer}{\end{eqnarray}}

\newcommand{\onethird}{\frac{\mbox{\small 1}}{\mbox{\small 3}}}

\newcommand{\tinyonehalf}{\frac{\mbox{\tiny 1}}{\mbox{\tiny 2}}}
\def\ket#1{\mbox{$\displaystyle\vert\,#1\,\rangle$}}

\begin{document}
\title{Helium-Three in Aerogel}
\author{W.P. Halperin and J.A. Sauls}
\affiliation{Department of Physics and Astronomy,
             Northwestern University, Evanston, Illinois 60208}
\date{Version \today}
\begin{abstract}
Liquid $^3$He confined in silica aerogel provides us with a
unique system to study the effects of quenched disorder on the properties
of a strongly correlated quantum liquid. The superfluid phases display interplay
between disorder and complex symmetry-breaking.
\end{abstract}
\maketitle

\subsection*{Introduction}

\begin{figure}[h]
\centerline{\epsfxsize0.65\hsize\epsffile{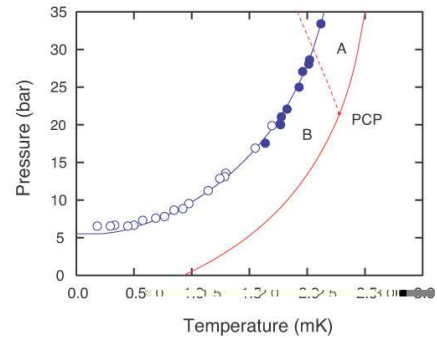}}
\begin{minipage}{0.9\hsize}
\caption{\label{PT-diagram}
Pressure vs. temperature phase diagram. The superfluid transition line for
pure $^{3}$He is the smooth red curve extending from $P =0-35$ bar. The A-B
transition for bulk $^3$He is shown as a dashed red line.
Shown here are the transitions for superfluid $^3$He in 98\% aerogel
from Cornell \cite{mat97} (open
circles) and Northwestern \cite{ger02} (solid circles).
There are modest variations in this phase diagram from one aerogel sample
to another at the same porosity.
The solid blue curve
is a theoretical calculation of the transition based on
the scattering model \cite{thu98,sau03}.
}
\end{minipage}
\end{figure}

The discovery of the superfluid phases of $^3$He led
to experimental and theoretical developments that have found deep
influence on many aspects of condensed matter physics. The original
observations were acknowledged with Nobel prizes
in 1996 to Douglas Osheroff, Robert Richardson and David Lee \cite{osh72},
and this past year for the
theoretical work \cite{leg72} of Anthony Leggett that developed
hand-in-hand with the early experimental investigations of these
phases. The transition from normal to superfluid is a second order
thermodynamic transition that was detected because of a discontinuous
jump in heat capacity. The line of phase transitions that marks the
onset of superfluidity is shown as the continuous red curve extending
from $P =0-35$ bar in Fig. \ref{PT-diagram}. Contrast this curve with the
blue data for 'dirty' superfluid $^3$He, showing that the effects of
impurities are to suppress the transitions and to create a zero temperature
critical pressure.

At relatively high
temperatures, the normal state of liquid $^3$He is well described by Landau's
Fermi liquid theory, formulated in terms low-lying excitations, called
``quasiparticles'', which are composite states of $^{3}$He atoms with spin
$\tinyonehalf$ and fermion number $1$ \cite{baym91}. Strong interactions also
lead to Bosonic excitations, e.g. phonons and spin-waves, but the fermionic
excitations dominate many of the low-temperature thermodynamic and transport
properties, including the specific heat, thermal conductivity, magnetization
and diffusion coefficient. At very low temperatures liquid $^3$He exhibits a
phase transition from a classic Fermi liquid state to a unique superfluid that
is a paradigm for many newly discovered ``unconventional'' superconductors
\cite{mineev99} in which one or more symmetries of the normal Fermi liquid
state (e.g. rotations, time-inversion, etc.) are spontaneously broken in
conjunction with the broken $U(1)$ gauge symmetry that is characteristic of
superconductivity. The ordered phases of pure superfluid $^3$He
are summarized in a ``sidebar''.

We can often obtain important information about new states of matter by
examining how their properties are modified by external influences. For
example, the effects of impurities and surfaces have played an important
role in revealing basic properties of the high temperature
superconductors, such as the nature of the order parameter
\cite{van95,lof01}, and they provide the key elements in applied areas of
superconductivity where flux pinning and critical currents are important.

Usually impurities and defects are unavoidable. However, in contrast to
metals, liquid $^{3}$He naturally expels impurities, even isotopic impurities,
which makes it the purest and most homogeneous condensed matter system, and one that,
until recently, could not be perturbed in a way that is typical of
superconductors where chemical impurities can be easily inserted into the
structure. In fact, some superconducting materials,
have chemical and physical imperfections which mask their intrinsic behavior.

In an earlier Physics Today article Chan et al. \cite{cha96} describe the
properties of superfluid \Hefour\ inside aerogel, notably the non-universal
critical behavior of superfluid $^{4}$He and the change of the phase diagram
for mixtures \Hefour\ and $^{3}$He. At that time the observation of
superfluidity of $^{3}$He in aerogel had just been made; more extensive work
has followed and is the subject of this article.

\begin{widetext}
\detail{
\noindent \textbf{Phases of Pure Superfluid $^{3}$He}\\

The superfluid phases of $^{3}$He are Bardeen-Cooper-Schrieffer condensates of
p-wave ($L=1$) Cooper pairs with orbital wave functions that are linear
superpositions of the p-wave states: $\Psi_{1,1}(\vr)=(x+iy)/\sqrt{2}$,
$\Psi_{1,-1}(\vr)=(x-iy)/\sqrt{2}$, $\Psi_{1,0}(\vr)=z$. The Pauli exclusion
principle then requires that these pairs form nuclear spin-triplet states ($S=1$).
There are three superfluid phases of pure $^3$He corresponding to different
realizations of the p-wave, spin-triplet manifold. Two phases, the A and B phases,
are indicated in the phase diagram in Fig. \ref{PT-diagram}.
A third phase, called the A${_1}$
phase, develops in a narrow region near T$_{c}$ in an applied magnetic field. All
three phases are characterized by their nuclear spin structure and
correspond to different superpositions of p-wave, spin-triplet states. The A${_1}$
phase is the spin-polarized state,
$\ket{A_1}=\Psi_{1,1}(\vr)\ket{\uparrow\uparrow}$, and has the highest transition
temperature in a magnetic field. The A phase is a superposition with equal
amplitudes for the oppositely polarized spin-triplet states (referred to as ``equal
spin pairing''), $\ket{A}=\Psi_{1,1}(\vr)\left(\ket{\uparrow\uparrow}+
\ket{\downarrow\downarrow}\right)/\sqrt{2}$. The A phase survives in large magnetic
fields without destroying Cooper pairs by conversion of
$\ket{\downarrow\downarrow}$ pairs into $\ket{\uparrow\uparrow}$ pairs in order to
accommodate the nuclear Zeeman energy. The B phase, which is the stable state over
most of the phase diagram in zero magnetic field, is a superposition of all three
triplet spin states:
$\ket{B}
=\Psi_{1,-1}(\vr)\ket{\uparrow\uparrow}
+\Psi_{1,1}(\vr)\ket{\downarrow\downarrow}
+\Psi_{1,0}(\vr)\ket{\uparrow\downarrow+\downarrow\uparrow}$ One of the key
signatures of a B-like phase is the reduction of the nuclear magnetic
susceptibility resulting from the $\ket{\uparrow\downarrow+\downarrow\uparrow}$
pairs. The B phase is suppressed in large magnetic fields when the Zeeman energy is
comparable to the binding energy of the
$\ket{\uparrow\downarrow+\downarrow\uparrow}$ pairs. The transition from A to B phases
is first order and is accompanied by a latent heat. This transition exhibits
supercooling, but no superheating. The point in the phase diagram where all three
phases are degenerate is called the polycritical point (PCP). This singular point
is destroyed by application of a magnetic field which opens up regions of stability
of both the $A$ and $A_1$ phases over the full pressure range.
Many of these features are fundamentally altered in the presence of disorder.
}
\end{widetext}
\bigskip

\subsection*{Aerogel}

\begin{figure}[!]
\centerline{\epsfxsize1.0\hsize\epsffile{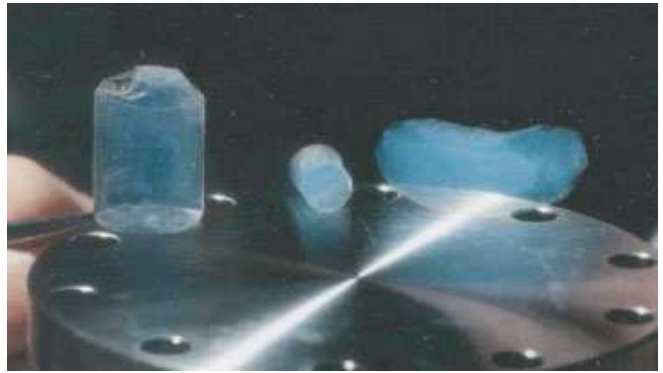}}
\begin{minipage}{0.9\hsize}
\caption{\label{photo}\small
Photograph of three silica aerogel samples with porosities 95\%, 98\%, and
99\% from left to right, placed on the top of a five inch diameter high pressure
autoclave used to supercritically dry the samples. The light blue color is
caused by Rayleigh scattering.
}
\end{minipage}
\end{figure}

In the past decade it was found \cite{por95,spr95} that impurities can be
introduced into $^3$He by impregnating the liquid into the open structure of silica
aerogel. These are extremely porous, low-density materials (as can be seen in Fig.
\ref{photo}), with porosities up to 99.5\% by volume, formed as a dilute network of
thin silica strands having a typical thicknesses of 3-5 nm \cite{fri88a}. Aerogels
are fascinating materials that have found practical applications, e.g. as Cerenkov
counters in particle physics and as light weight transparent thermal insulation.
Figure \ref{photo} is a photograph of three silica aerogels with widely ranging
porosities. The aerogels are transparent, clearly
evident for the most porous sample on the right. One can infer that the structure is
homogeneous on length scales of order the wavelength of visible light.

\begin{figure}[h]
\centerline{\epsfxsize0.8\hsize\epsffile{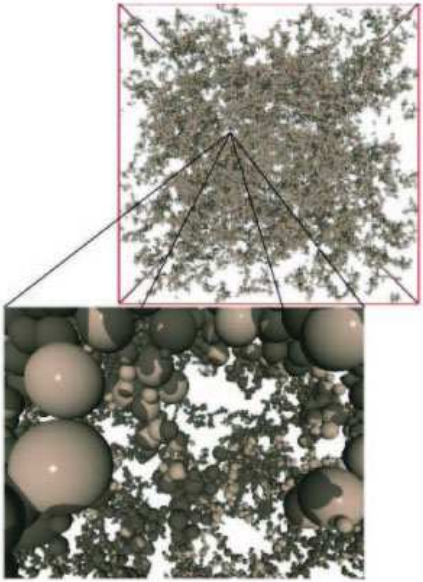}}
\begin{minipage}{0.93\hsize}
\caption{\label{DLCA}\small
Perspective view of a $98$\% porosity gel structure grown by numerical
simulation \cite{haa00} for a $600\times 600\times 600\,\text{nm}{^3}$ volume
beginning with a random suspension of particles having a log-normal distribution and
mean diameter of $3\,\text{nm}$. The particles assemble in strands that are spatially
correlated over distances of order $30 \,\text{nm}$. The calculated geometric mean
free path in this gel was found to be $200\,\text{nm}$.
}
\end{minipage}
\end{figure}

Silica aerogel is formed from a synthesis of silica clusters
approximately $\delta=3\,\text{nm}$ in diameter. Gelation is performed from
tetramethylorthosilicate and the clusters aggregate to generate the strands that
form the final gel
structure. The wet gel is dried at a supercritical pressure in a high pressure
autoclave to avoid collapse of the microstructure from capillary forces at the
liquid-gas interface. The resulting material is air stable and
hydrophobic. Small angle X-ray measurements \cite{cha96,por99} indicate that
there are fractal correlations characteristic of the process of
diffusion-limited cluster aggregation over a decade or more in wavevector.
Density correlations are observed to onset at $q_a^{-1}\simeq 10-30$ nm
in the structure factor. The aerogel correlation length,
$\xi_a=\pi/q_a$, is identified as the typical distance between silica strands or
clusters, $\xi_a\approx 30-100\,\text{nm}$. At
longer length scales the aerogel particle-particle correlations are random.
These conclusions are supported by numerical simulations of a 98\% porosity gel
structure shown in Fig. \ref{DLCA} \cite{haa00}. In the simulation an ensemble
of particles, which is initially randomly distributed, executes Brownian motion
until the particles aggregate by contact with one another. For the
simulated structure shown in Fig. \ref{DLCA} a geometric mean free path of
$200\,\text{nm}$ was obtained as the average length of a straight line
trajectory terminating on aerogel particles. This geometric mean free path is
what one expects for the transport mean free path resulting from elastic
scattering of $^{3}$He quasiparticles moving at constant speed through the pore
space. The geometric mean free path is indeed comparable to the transport mean
free paths, $\lambda\approx 130-180\,\text{nm}$, that have been obtained from
analyses of transport measurements on liquid $^3$He (spin diffusion, acoustic
attenuation, and thermal conductivity) performed in a 98\% aerogel. The close
comparison of these length scales provides support for the application of
scattering theory to describe the effects of aerogel on the superfluid phases
of $^3$He.

\subsection*{$^3$He with Quenched Disorder}

When impregnated with $^3$He, the aerogel is found to have a dramatic effect on the
properties of liquid $^3$He. The superfluid transition temperature, $T_c$, is
suppressed well below the bulk value for all pressures \cite{por95,spr95}, and there
is a zero-temperature, ``quantum'' phase transition \cite{mat97} at $p_c\simeq
6\,\mbox{bar}$, for a 98\% porous aerogel, separating a disordered normal
Fermi-liquid phase from a superfluid
phase with very different properties than that of pure $^3$He. Quenched disorder
leads to new physical behavior in a quantum liquid with
complex symmetry breaking. Its detailed study may help us
better understand the pure phases of $^{3}$He as well as strongly correlated
electronic materials with unconventional pairing.

\begin{figure}[h]
\centerline{\epsfxsize0.65\hsize\epsffile{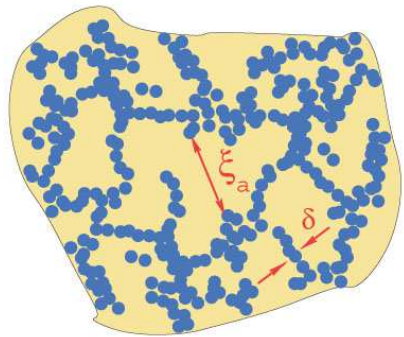}}
\begin{minipage}{0.9\hsize}
\caption{\label{aerogel_sketch}
A sketch of silica aerogel showing low-density regions containing $^3$He (yellow)
threaded by higher density strands and aggregates of silica (blue). Two principal
length scales are indicated: the typical size of the aerogel strands, $\delta\simeq
3\,\text{nm}$, and the aerogel correlation length, $\xi_a\simeq 30\,\text{nm}$,
identified as the average inter-strand distance.
}
\end{minipage}
\end{figure}

The two length scales, $\delta$ and $\xi_a$, characteristic of the structure of aerogel
are identified in the sketch shown in Fig. \ref{aerogel_sketch}.
These length scales are much larger than the Fermi wavelength of $^3$He quasiparticles,
$\approx 0.1\,\text{nm}$.
Consequently, to an excellent approximation, the basic properties
of the normal Fermi liquid in the open volume, such as the density,
effective mass and quasiparticle interactions, are essentially unaffected by the
aerogel. This is not the case for the superfluid phase. Here the important length is
the coherence length of
pure superfluid $^3$He (the size of a Cooper pair), $\xi\approx 80$ nm at low
pressure, which is comparable to the aerogel correlation length, $\xi_a$. One expects
significant effects on the phase diagram and structure of the Cooper pair
condensates in such a confined geometry.

In pure, bulk $^3$He inelastic binary collisions between quasiparticles limits the
transport of heat, momentum and magnetization. Elastic scattering is absent, except
at ultra-low temperatures when boundary scattering from the containing walls limits
ballistic propagation of quasiparticles. Aerogel changes this situation. At
temperatures below $T^{\star}\approx 10\,\text{mK}$ \textit{elastic} scattering of
quasiparticles by the aerogel dominates \textit{inelastic} quasiparticle-quasiparticle
collisions \cite{rai98}. Elastic scattering limits the mean free path of normal
$^3$He quasiparticles to $\lambda\simeq 130-180\,\text{nm}$ for aerogels with $98\%$
porosity. Thus, the low-temperature limits for the transport coefficients are
determined by scattering from the aerogel. Consequently, transport processes
in $^{3}$He are similar to those typical of metals at low temperatures.
Experimental measurements for
$T\ll T^{\star}$ provide a direct determination of the transport mean free path.
Once it is determined we can make quantitative predictions for the thermodynamic and
transport properties of the superfluid phase of $^3$He in aerogel, and test the
scattering theory.

\subsection*{Normal-state transport}

In the normal state of pure liquid $^3$He, the quasiparticle lifetime, or mean free
path, is determined by the
inelastic collision rate between quasiparticles \cite{abr58},
\be
\frac{1}{\tau_{\text{in}}}= \langle W \rangle\frac{\left(k_B T\right)^2}{E_f}\propto
T^2
\,,
\ee
where $\langle W \rangle$ is the transition probability for quasiparticle scattering
on the Fermi surface and $E_f$ is the Fermi energy \cite{baym91}. For $^3$He in
aerogel, binary collisions dominate at relatively high temperatures, while
elastic scattering of quasiparticles from the aerogel strands leads to a
temperature-independent scattering rate at low temperatures ($T < T^{\star}$).
The latter is determined by the transport mean free path, $\lambda$, where
\be\label{taul}
\frac{1}{\tau_{\text{el}}} = \left(v_f/\lambda\right)
\,,
\ee
and $v_f$ is the Fermi velocity.
A good illustration of the cross-over is provided by the transport of magnetization
in the hydrodynamic limit, given by the spin current density,
\begin{equation}
\vj_{\text{M}} = -\diff\,\grad\,M
\,,
\end{equation}
where $M$ is the local magnetization and $\diff$ is the spin diffusion coefficient.

\begin{figure}[h]
\centerline{\epsfxsize0.75\hsize\epsffile{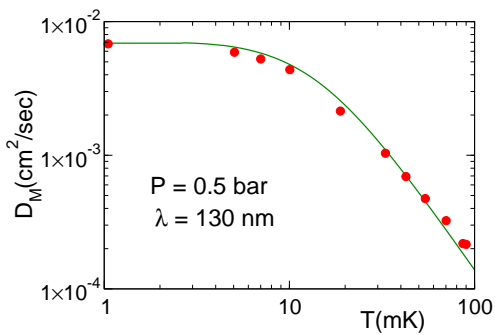}}
\begin{minipage}{0.9\hsize}
\caption{\label{Spin_Diffusion-Grenoble}
Comparison of the theory to experimental data taken at Grenoble \cite{col02}.
The inelastic
scattering rate is fit to the high-temperature data. The elastic transport mean
free path obtained from the fit is $\lambda=130$ nm.}
\end{minipage}
\end{figure}

The diffusion coefficient can be calculated from Landau's kinetic
theory of quasiparticles, with the collision integral determined by
both elastic scattering from the aerogel medium and inelastic collisions
between quasiparticles. The general solution has the form,
\be\label{DM}
\diff=\onethird v_f^2(1 + F_0^a)\taud
\,,
\ee
in the hydrodynamic limit, $\omega_{\text{\tiny L}}\ll\taud^{-1}$, where
$\omega_{\text{\tiny L}}=\gamma B$ is the Larmor frequency, $\taud^{-1}$ is the
collision rate that limits the transport of magnetization
and $F_0^a$ is the exchange interaction for liquid \He. The diffusion
coefficient has been calculated from an exact solution of the Landau-Boltzmann
transport equation including both scattering channels \cite{sha00}.

Measurements of the spin-diffusion coefficient for $^3$He in $98\%$ aerogel performed
in Grenoble \cite{col02} are shown in Fig.
\ref{Spin_Diffusion-Grenoble} for $P =0.5\,\text{bar}$. The spin-diffusion
coefficient decreases as $\diff\propto T^{-2}$ at high temperatures and coincides
with measurements of the spin-diffusion coefficient for bulk $^3$He \cite{and62}.
At low temperature there is a cross-over to the elastic scattering regime
determined by the aerogel. The mean-free path is found to be $\lambda=130$ nm
for this 98\% aerogel.

\subsection*{Aerogel Scattering}

The fact that the superfluid coherence length is much larger than the silica strand
dimensions and comparable to or larger than the aerogel correlation length,
at least at lower pressures, suggested that a theory based on atomic scale
scattering centers might provide an adequate description of the effects of the
aerogel on the properties of superfluid $^3$He. A theoretical approach based
on scattering by point impurities, analogous to the Abrikosov-Gorkov theory
of disorder in superconductors \cite{abr61}, was developed by Thuneberg {\it
et al.} \cite{thu98} in an effort to account for the observed behavior.
Elastic scattering by impurities and defects
results in diffusive transport in the normal Fermi liquid state, and to
substantial corrections to the transition temperature, order parameter, quasiparticle
excitation spectrum, superfluid density, magnetization, etc. In its simplest
form the theory assumes that elastic scattering of $^3$He quasiparticles by
the silica aerogel is isotropic and homogeneous; this is referred to as the
homogeneous, isotropic, scattering model (HISM). Extensions of the theory
\cite{thu98,sau03} to include inhomogeneities of the scattering medium,
referred to as inhomogeneous, isotropic scattering models (IISM), are
required when aerogel structural correlations are comparable to or
larger than the coherence length, as is the situation at higher pressures.

\begin{figure}[t]
\centerline{\epsfxsize 0.8\hsize\epsffile{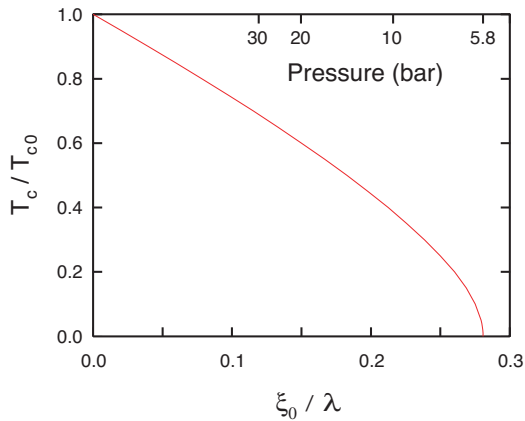}}
\begin{minipage}{0.9\hsize}
\caption{\label{AG}\small
The reduction of the transition temperature of an unconventional superconductor with
coherence length, $\xi_0$, owing to elastic scattering of Fermi quasiparticles with
mean free path $\lambda$. The pressure scale was calculated for $\lambda=140$ nm.}
\end{minipage}
\end{figure}

Elastic scattering of quasiparticles is deleterious to an unconventional superfluid,
like $^{3}$He. The fragility of non-{\it s}-wave pairing holds for superfluid
$^{3}$He as well as for a number of superconducting compounds, including the
heavy fermion superconductor UPt$_3$ ({\it f-}wave), the copper oxide superconductors
({\it d-}wave), Sr$_2$RuO$_4$ ({\it p-}wave) and possibly several organic
superconductors. Scattering from impurities reduces the coherence between pairs of
quasiparticles that bind to form Cooper pairs, thus reducing the
transition temperature and suppressing the magnitude of the order parameter. The
Abrikosov-Gorkov theory, originally developed for magnetic scattering in conventional
isotropic superconductors, is easily generalized to describe Cooper pair breaking in
unconventional (non-{\it s-}wave) superfluids for magnetic or non-magnetic impurity
scattering.

In the HISM the pair breaking parameter that determines the suppression of the
transition temperature is the ratio of the pure superfluid coherence length to the
transport mean free path, $\xi_0/\lambda$. In fact the critical
pressure ($p_c\simeq 6$ bar) shown in Fig. \ref{PT-diagram} for a 98\% porous aerogel
corresponds to the critical pair-breaking parameter shown in Fig. \ref{AG}, i.e.
$\xi_0(p_c)/\lambda = 0.28$. For $^{3}$He in aerogel the limiting mean free path is a
constant, independent of pressure, fixed by the aerogel structure. But the coherence
length at zero temperature, $\xi_0=\hbar v_f/2\pi k_B T_{c0}$,
varies from $77$ nm at low pressure to $16$ nm at the
highest pressures near the melting curve. For a mean free path of $140$ nm, which
is close to the value obtained from the spin-diffusion measurements
for the 98\% aerogel used by the Grenoble group, we can
account for the observed critical pressure, and implicitly from the theory we obtain
the pressure scale shown on the upper axis of Fig. \ref{AG}. Rotating Fig. \ref{AG}
clockwise by $90$ degrees shows the qualitative agreement with the experimental phase
diagram for $^3$He in aerogel shown in Fig. \ref{PT-diagram}. The HISM provides a
reasonable description of the dirty superfluid at low
pressures; it accounts semi-quantitatively for the reduction of $T_c$, including the
critical pressure, $p_c$, and the pair-breaking suppression of the order parameter
\cite{thu98,rai98}. However, the HISM under estimates the transition temperature
at higher pressures and higher porosities where the pair size is
comparable to, or smaller than, the typical distance between aerogel strands. This
failure of the HISM is most evident in the pressure dependence of $T_c$ \cite{thu98,law00}.

The qualitative picture of the correlated aerogel
is a random distribution of low density regions, `voids', with a typical
dimension of $\xi_a$. These low-density regions are available for formation of the condensate
at higher temperatures.
In the limit $\xi_0\ll\xi_a$, the suppression of the superfluid transition is
determined by a new pair-breaking parameter, proportional to $(\xi_0/\xi_a)^2$.
In the opposite limit, when the pair size is
much larger than $\xi_a$, the aerogel is effectively homogeneous on the scale of the
pairs and pair-breaking results from homogeneous scattering defined by the pair-breaking
parameter $(\xi_0/\lambda)$.
This latter limit is achieved at low pressures.
Theoretical analyses which include the effects of
aerogel correlations \cite{thu98,han03,sau03} provide a quantitative description of
the phase diagram, as well as the order parameter, excitation spectrum and transport
properties of $^3$He in aerogel over the whole pressure range.

\subsection*{Superfluidity}

\begin{figure}[h]
\centerline{\epsfxsize 0.8\hsize\epsffile{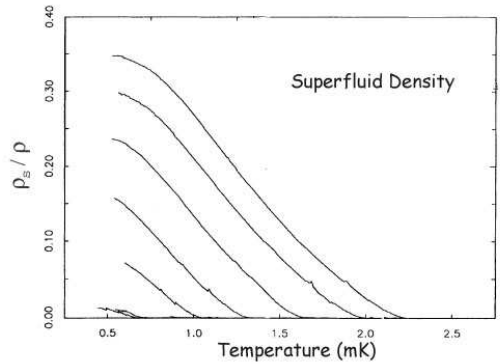}}
\begin{minipage}{0.9\hsize}
\caption{\label{rhos}\small
The superfluid fraction as determined from torsional oscillator measurements from the
Cornell group \cite{por95}. Measurements at pressures of (from lowest to highest)
$3.4$, $5.0$, $7.0$, $10$, $15$, $25$ bar. The superfluid fraction is smaller
than that of pure $^3$He which approaches $\rho_s/\rho=1$ as $T\rightarrow 0$.}
\end{minipage}
\end{figure}

A powerful technique that is sensitive to the onset of superfluidity makes use
of a high-Q torsional oscillator. The torsion rod is attached perpendicular to a
disk containing the helium sample. Using this approach, Porto and Parpia at Cornell
\cite{por95} found that the period of the oscillator shifted with an abrupt onset at
specific temperatures depending on the pressure.
The method relies on the fact that the normal fluid is viscously
clamped to the porous structure; but since the superfluid has zero viscosity it does
not contribute to the moment of inertia. At the onset of superfluidity there is a
sharp decrease of oscillation period. The loss of inertia is quantitatively
interpreted in terms of the superfluid density, shown in Fig. \ref{rhos}.
%
%
In contrast with pure superfluid $^3$He the superfluid fraction is much less than unity,
a direct reflection of the pair-breaking effect of scattering off the aerogel. Shortly
after these measurements were performed the group at Northwestern \cite{spr95} observed
the sharp onset of frequency shifts in the nuclear magnetic resonance (NMR) spectrum,
qualitatively similar to those associated with the superfluid phase in pure $^{3}$He.
But in the case of $^3$He in aerogel the shifts were significantly reduced in magnitude.
The reduction of NMR frequency shifts and superfluid fraction compared to pure $^3$He
provides a consistent picture for the reduction of the magnitude of the order parameter.
This fact is directly confirmed in measurements of the heat capacity.

\begin{figure}[!]
\centerline{\epsfxsize 0.8\hsize\epsffile{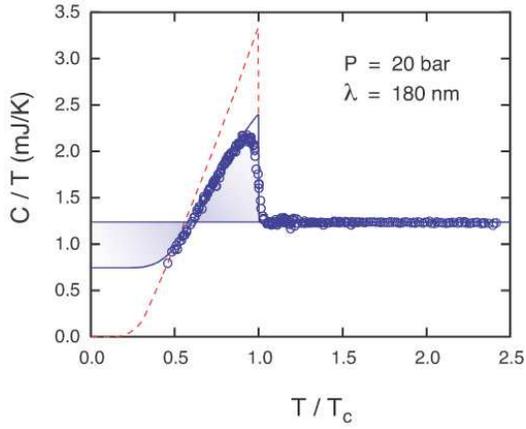}}
\begin{minipage}{0.9\hsize}
\caption{\label{CvsT}\small
The heat capacity of $^3$He in a 98\% porous silica aerogel. The measurements, performed
at Northwestern University, show the transition to the superfluid state as a jump in
the heat capacity which is reduced by a factor of 0.55 compared to that of
pure superfluid $^3$He at the same
pressure of 20 bar (dashed red curve).}
\end{minipage}
\end{figure}

Heat capacity experiments \cite{cho04}, shown in Fig. \ref{CvsT} for a pressure of
$20$ bar, demonstrate that the superfluid transition has a discontinuity similar
to that of a BCS superconductor. For pure superfluid $^3$He the size of the
discontinuity is larger than the predicted BCS result of $\Delta C/C = 1.43$.
Strong coupling effects are responsible for specific heat jumps as large as
$\Delta C/C\simeq 2.0$ near melting pressure. Direct comparison between pure
superfluid $^3$He (dashed red curve) \cite{gre86} and that of superfluid $^{3}$He
in aerogel (shown in blue) indicates that the heat capacity discontinuity is
substantially reduced, by a factor of 0.55 at a pressure of $20\,\text{bar}$, due
to scattering by the aerogel. The reduction is even below that expected for the
weak-coupling limit of a clean BCS superconductor. This is a consequence of two
factors: (i) the decrease in the transition temperature leads to a reduction of
strong coupling effects, proportional to T$_c$, and (ii) the pair-breaking effect
increases the free energy and reduces the magnitude of the order parameter even
in the weak coupling approximation \cite{thu98}.

\subsection*{Thermal Conductivity}

Heat transport by diffusion in pure superfluid $^3$He is masked by the
propagation of heat via hydrodynamic mode in which
normal and superfluid components move counter to one another
\cite{whe75}. The presence of aerogel strongly reduces the mean free path
for quasiparticle diffusion, and effectively clamps the hydrodynamic motion
of the normal component in ``dirty'' superfluid $^3$He.
Hydrodynamic heat flow is suppressed and heat transport is
determined by quasiparticle diffusion.

\begin{figure}[!]
\centerline{\epsfxsize 0.8\hsize\epsffile{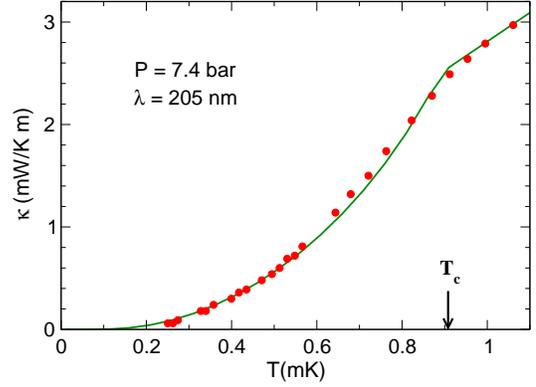}}
\begin{minipage}{0.9\hsize}
\caption{\label{Thermal_Conductivity-Lancaster}\small
Thermal conductivity data from the Lancaster group \cite{fis01} (circles) at
$P =7.4\,\mbox{bar}$ is compared with theory (solid
curve). The theoretical calculation assumes a B-phase order parameter. A mean free
path of $\lambda=205\,\text{nm}$ is obtained from the slope of the normal-state thermal
conductivity. The theoretical results, including the value of $T_c$ for this aerogel,
are based on the same mean free path.
}
\end{minipage}
\end{figure}

Measurements of the thermal conductivity of $^3$He in 98\% aerogel carried out at
the University of Lancaster are shown in Fig. \ref{Thermal_Conductivity-Lancaster}.
The signature of the onset of superfluidity in thermal conductivity is the change
in slope at $T_c$.

Scattering by the aerogel matrix leads to pair-breaking and the formation of a
spectrum of low-energy quasiparticle states below the gap. In general the excitation
spectrum depends upon the symmetry of the order parameter, as well as the scattering
cross-section and mean-free path. The suppression of the thermal conductivity
shown in Fig. \ref{Thermal_Conductivity-Lancaster} below that of the normal state is
characteristic of elastic scattering and the suppression of the density of states
for low-energy excitations. The measurements of the thermal conductivity of $^3$He
in $98\,\%$ aerogel at low pressures \cite{fis01} are in good agreement with
theoretical calculations based on either the B- or A-phase order parameters. At
higher pressures, where pair-breaking effects are weaker, significant differences in
the thermal conductivity for these two phases are predicted \cite{sha01}.

\subsection*{Gapless Superfluid}

In the strong scattering limit of the HISM a band of gapless excitations forms,
{\sl centered at the Fermi level}, with energies $\varepsilon\le\gamma\approx
0.67\Delta\sqrt{\xi_0/\lambda}$. This band is relatively insensitive to the
symmetry of the order parameter, but depends on its magnitude, $\Delta$. The
constant density of states for low-energy excitations, $\varepsilon\ll\gamma$,
gives rise to a linear $T$-dependence of both the thermal conductivity and
specific heat deep in the superfluid phase at very low temperatures,
$k_BT\ll\gamma$.

Heat transport measurements by the low-temperature group at the University of
Lancaster\cite{fis03} have shown that the low-temperature limit of the thermal
conductivity of superfluid $^3$He is linear in temperature, consistent
with there being a significant density of gapless Fermion excitations near the Fermi
level.

The third law of thermodynamics requires that the entropy of both the normal
and superfluid states vanish at zero temperature. For a second-order
transition the equality of the entropy for both normal and superfluid phases
at T$_c$ requires that the shaded areas in Fig. \ref{CvsT} to be equal. If we
rule out an unphysical, non-monotonic, temperature dependence of the heat
capacity at low temperatures, the equal-area constraint requires that there be
a non-zero intercept for $C/T$ at $T=0$, as shown in Fig. \ref{CvsT}.
Consequently, the heat capacity must be linear in $T$ at low temperatures. The
intercept is directly proportional to the density of impurity-induced
quasiparticle states at the Fermi level. For comparison, the B phase of pure
superfluid $^3$He is fully gapped over the
entire Fermi surface. The evidence is compelling from both of these thermal
experiments that liquid $^3$He in aerogel is a gapless superfluid.

\subsection*{Metastability}

While it is widely accepted that a superfluid phase can exist in a sufficiently
dilute aerogel, the precise nature of the superfluid phase is not fully resolved.
Fundamental questions have been raised about the nucleation, stability and symmetry
of phases that may be stabilized within the aerogel structure
\cite{ger01,ger02,bar00}. Theoretical calculations show that homogeneous and
isotropic impurity scattering stabilizes the isotropic state (B phase) relative to
the axial state (A phase) \cite{thu98,bar02}.
This has been confirmed for 98\% aerogels \cite{ger02c}.

\begin{figure}[!]
\centerline{\epsfxsize 0.8\hsize\epsffile{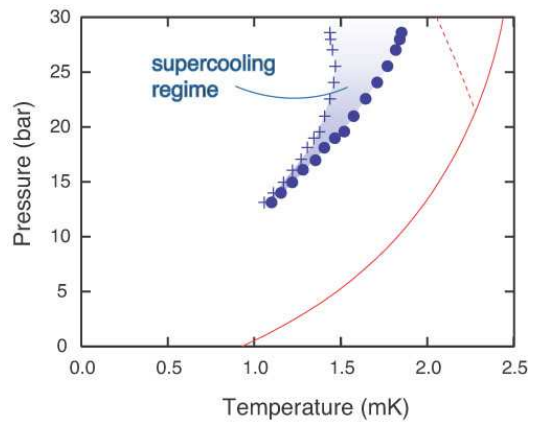}}
\begin{minipage}{0.9\hsize}
\caption{\label{PT-metastable}\small
The supercooled region of superfluid $^3$He-A in a 98\% porous silica aerogel is
shown in blue. The measurements are from Cornell \cite{naz04}.
}
\end{minipage}
\end{figure}

Nevertheless, early measurements of susceptibility and NMR frequency shifts
indicated that an ESP state was observed, like the A phase in pure $^3$He. NMR
experiments performed by the Stanford group have amplified on these earlier
findings. They discovered \cite{bar00} that on cooling there is a transition
between two superfluid phases which is highly hysteretic and that the lower
temperature phase has a decreased susceptibility, like the B phase (shown in Fig.
\ref{Susceptibility-Stanford}), and that the higher temperature superfluid phase
has a constant susceptibility like the A phase. This metastable A-phase region is
shown shaded blue in Fig. \ref{PT-metastable}. These conclusions were based on
both observations of nuclear magnetic susceptibility obtained from the integral
of the NMR spectrum, as well as from abrupt NMR spectral shifts, which can be
negative only in the case of an A phase. To account for hysteresis either the A phase
supercools or the B phase superheats. Substantial effort has been directed
to resolving which effect is dominant and what is the correct identification of
these two phases. This includes work from laboratories at Stanford, Cornell,
Northwestern, Lancaster, Osaka City University,
University of Florida and the Kapitza Institute in Moscow.

The region of the phase diagram in question is shown in Fig. \ref{PT-metastable},
taken from the Cornell measurements. Using transverse ultrasound the Northwestern
group \cite{ger02} showed that there is at most a tiny equilibrium sliver ($\Delta
T\simeq 20\mu$K) of A phase just below T$_c$, and that there is no PCP for $^3$He in
98\% aerogel. They concluded that a B-like phase is stable over essentially all of the
equilibrium phase diagram in zero field. How is it possible to have a supercooled
A phase in zero magnetic field as shown in Fig. \ref{PT-metastable}, if it is not
thermodynamically stable at higher temperature? And if it is stable, why is this phase
restricted to such a narrow interval below $T_c$? This problem is not yet resolved and
likely involves physical ideas for which there are no obvious parallels in pure
superfluid $^3$He. Recently, the Stanford and Cornell experimenters found that the
shaded region in Fig. \ref{PT-metastable} can, under certain conditions, sustain
quasi-stable mixtures of A and B phases. Possibly the interfaces between the A and
B-phase domains are pinned by the aerogel structure, reminiscent of impurity pinning
of domain walls between magnetic phases or flux phases
in other condensed matter systems.

\subsection*{Magnetic Susceptibility}

The nuclear magnetization of $^3$He played a central role in the identification of
the spin-structure of the phases of pure superfluid $^3$He. For equal-spin-pairing
(ESP) states, like the A phase,
the magnetization is unchanged relative to that of the normal state since
relative populations of the $\ket{\downarrow\downarrow}$ and $\ket{\uparrow\uparrow}$
pairs can be shifted without pair-breaking to accommodate the nuclear Zeeman energy.
However for non-ESP phases like the B phase, the nuclear spin susceptibility is
reduced by the formation of $\ket{\uparrow\downarrow+\downarrow\uparrow}$ pairs.

\begin{figure}[!]
\centerline{\epsfxsize 0.8\hsize\epsffile{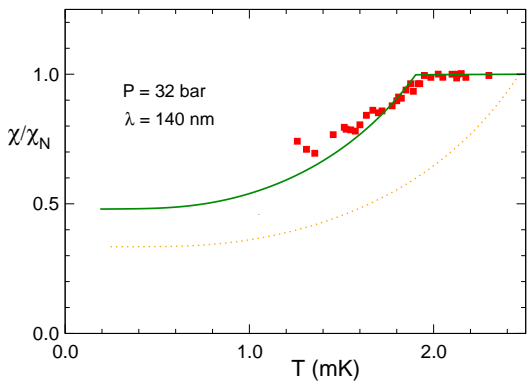}}
\begin{minipage}{0.9\hsize}
\caption{\label{Susceptibility-Stanford}\small
Experimental data for the magnetic susceptibility of $^3$He in 98\% aerogel
at $P =32.0\,\text{bar}$ \cite{bar00}. The solid green curve is the
theoretical result for a ``dirty B phase'' with a
mean-free path of $\lambda=140\,\text{nm}$ and
a correlation length $\xi_a=40$ nm.
The susceptibility of pure
$^3$He-B is also shown for comparison (orange dotted curve).
}
\end{minipage}
\end{figure}

Changes in the nuclear magnetization of superfluid $^3$He in aerogel are
determined by competing effects of pairing correlations of quasiparticles
with $S_z=0$, and pair-breaking induced by scattering from the aerogel
structure. In addition to the suppression of $T_c$ by scattering from the
aerogel, the magnitude of the susceptibility, particularly at low
temperatures, is sensitive to the polarizability of the sub-gap excitations.
Measurements of a B-like susceptibility in $^3$He-aerogel have been reported
\cite{spr95,bar00}. Theoretical results for the susceptibility of superfluid
$^3$He in aerogel with a B-phase order parameter were obtained
\cite{sha01,min02} and compared with the experimental results. Figure
\ref{Susceptibility-Stanford} shows measurements of the susceptibility from
the Stanford group \cite{bar00}, as well as theoretical results for the
susceptibility of a B-phase order parameter with aerogel correlation effects
included within the effective pair-breaking model
\cite{sau03}. The aerogel strand correlations lead to both an
increase in $T_c$ and a decrease in the susceptibility due to
a reduction in the sub-gap polarizability at low
temperatures for the same mean free path.

\subsection*{New Directions}

\subsubsection*{Fundamentally new phases in $^3$He-aerogel?}

While there is support for the identification of the principal equilibrium
superfluid phase in 98\% aerogel as the dirty B phase, it is less certain that
the orbital symmetry of the metastable phase is an A-like phase. There is less
known about the energetics that governs the strong metastability of
this phase. Theoretical understanding of defect structures in pure
$^3$He give no \textit{a priori} reason to assume that the stable
or metastable phases of superfluid $^3$He should be simply related
to the stable bulk phases of pure $^3$He. To the contrary, surface
scattering and strong spatial variations of the order parameter
imposed by scattering from an inhomogeneous distribution of aerogel
strands and particles suggest that new phases, not realized in the
bulk of pure $^3$He, may be stabilized in aerogel. This is
certainly the case if the aerogel has significant orientational
correlations on the coherence length scale. Long range
orientational correlations of the silica strands scatter
quasiparticles anisotropically and lead to anisotropic pair
breaking. This implies the possibility of new phases reflecting the
locally broken rotational symmetry of the aerogel. A theoretical
suggestion for a new class of ``robust'' phases of $^3$He in
aerogel that are stabilized by accommodating anisotropy in the
scattering medium has been proposed \cite{fom03}. Whether or not
this phase or more complex structures are identified with the
superfluid phase(s) of $^3$He in extremely porous aerogels
($>99\%$) requires new theoretical and experimental developments.
Perhaps one of the more direct tests of these ideas is to impose
anisotropy on a macroscopic scale within the aerogel. This would
open up a new direction for the study of the orbital order
parameter of $^3$He, analogous to the Zeeman coupling to the spin
degrees of freedom of the order parameter. Indeed, such anisotropy
can stabilize a two-dimensional planar or axial phase or a
one-dimensional polar phase, depending on the nature of the induced
anisotropy. Systematic studies of anisotropy-induced stabilization
of new phases may also provide clues to the nature of the
metastable phase in isotropic aerogels. For example, it is also
possible that short-range anisotropic strand-strand correlations
provide a mechanism for the stabilization or metastability of an
A-like phase, even at pressures below the PCP of pure $^3$He.

\subsubsection*{Does $^3$He coating the aerogel strands lead to new physics?}

One of the early observations from NMR was the presence of a strong Curie
susceptibility at milli-Kelvin temperatures. The origin of this magnetization is
the localized $^3$He that forms one or two mono-layers of solid
$^3$He on the surface of the silica strands. This solid $^3$He can be removed by
preferentially coating the silica with a few monolayers of $^4$He. But, the
question arises as to whether or not the solid $^3$He spins play any significant
role in scattering $^3$He quasiparticles and the process of pairing breaking. It is
generally assumed, although direct evidence is sketchy, that for liquid $^3$He in
contact with solid $^3$He there is an indirect exchange
coupling between the $^3$He spins in the solid and the itinerant $^3$He
quasiparticles of the liquid phase. Is this coupling present, and if so what is the
sign and magnitude of the indirect exchange coupling?

Independent of the precise nature of the orbital state of superfluid $^3$He
near $T_c$, the Zeeman coupling of a magnetic field to the spin-triplet
Cooper pairs predicts a splitting of the phase transition in a magnetic field
and stabilization of pure $\ket{\uparrow\uparrow}$ Cooper pairs in a narrow
temperature interval that increases linearly with the field, $\Delta
T_c\approx (60\mu\text{K}/\text{T})\,B$. Indications of a solid-liquid
exchange interaction come from low-field measurements of the phase diagram.
For $^3$He in 98\% aerogel it was found that there is no change in $T_c$ and
no evidence of a splitting of the transition for fields up to $0.8$
Tesla \cite{ger02c}. Recent experiments at the University of Florida, in fields ranging
from $5$ to $15$ Tesla, clearly show this splitting, but one which is smaller
in the dirty superfluid compared to that in pure $^3$He.

\begin{figure}[!]
\centerline{\epsfxsize 0.8\hsize\epsffile{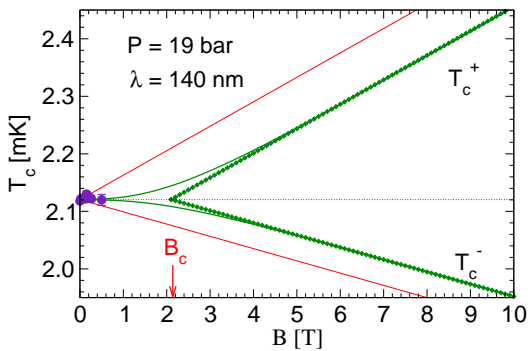}}
\begin{minipage}{0.9\hsize}
\caption{\label{A1-A2_splitting}\small
         Theoretical prediction for the nonlinear field evolution of the
         splitting of $T_c$ with magnetic
         field for \He\ in aerogel with a mean free path of
         $\lambda=140\,\mbox{nm}$, a correlation length of $\xi_a=50\,\text{nm}$
         and an anti-ferromagnetic exchange coupling of $J\simeq 0.1\,\text{mK}$
         (green lines). The splitting for \He-aerogel without
         liquid-solid exchange is indicated by the solid (red) lines.
         The extrapolation of the high-field splitting to $T_c(0)$ provides
         a direct measure of the exchange field, $B_c\approx 2\,\text{T}$ (dark green lines).
         For a ferromagnetic coupling the exchange field would have the opposite sign.
}
\end{minipage}
\end{figure}

If the splitting of the transition is suppressed in aerogel then a new mechanism
must be at work that competes with the interaction responsible for the splitting
of the transition in pure superfluid $^3$He. One possibility is the indirect
exchange coupling
with the solid $^3$He spins \cite{sau03,bar00c}. This coupling gives rise to an
additional term for
the splitting, linear in the field, which can support or compete with the
intrinsic mechanism responsible for the splitting in pure $^3$He. Indeed for an
anti-ferromagnetic exchange coupling of $J\approx 0.1\,\text{mK}$ per $^3$He
spin the low-field splitting, $B< 1\,\text{T}$, is suppressed. However, at
higher fields the magnetization of the solid $^3$He will saturate and the
splitting will appear for fields above the exchange field. Figure
\ref{A1-A2_splitting} shows the predicted evolution of the splitting at higher
fields. Note that by extrapolating the linear splitting at high fields to
$T_c(B=0)$ determines
the exchange field, $B_c$, thus, providing a test of the theoretical proposal that
indirect exchange between liquid and solid $^3$He in aerogel is present and
modifies the phases of superfluid $^3$He. The presence of an indirect
liquid-solid exchange coupling of this magnitude would open new directions for
studying solid $^3$He magnetism in reduced dimensions, as well as a range of
new low-field transport phenomena, in both the normal and superfluid phases,
associated with spin-dependent scattering from polarized solid $^3$He.

\subsubsection*{New phenomena of $^3$He in aerogel?}

The impregnation of $^3$He in a solid structure which also suppresses the
bulk transition provides possibilities for the study of
proximity effects, mesoscopic transport and Josephson effects based on interfaces
and weak links established between pure $^3$He and
$^3$He confined in aerogel. Initial studies of nucleation and possible proximity
effects between bulk superfluid $^3$He and $^3$He-aerogel have already opened up
new questions about the mechanism(s) for nucleation of phases, pinning of order
parameter structures and mechanisms for meta-stability of inhomogeneous states
of $^3$He in aerogel.

Finally, the impregnation of liquid $^3$He within an elastic solid with negligible
intrinsic dissipation provides a new arena for the
study of the collective mode dynamics of
superfluid phases. In addition to longitudinal acoustic waves, the solid aerogel
provides a mechanism to excite the confined liquid with a bulk transverse current
probe. This opens up the possibility of studying new mechanisms for acoustic
birefringence that reflect the underlying broken symmetries of the superfluid phases.
Will the ground state of superfluid $^3$He in aerogel exhibit spontaneous birefringence
associated with broken time-inversion symmetry, or broken chiral symmetry?
Will short-range anisotropic correlations of the aerogel lead to linear birefringence?
Can we tailor new aerogels with chiral properties and induce new types of broken symmetry
in the liquid? These are some of the many interesting and challenging directions to pursue
in future work.


\end{document}